\documentclass[preprint2]{aastex}

\shorttitle{Calcium-rich Gap Transients and the Intracluster Medium}
\shortauthors{Mulchaey et al.}

\begin{document}

\title{Calcium-rich Gap Transients: Solving the Calcium Conundrum in the Intracluster Medium}

\author{John S. Mulchaey, Mansi M. Kasliwal \altaffilmark{1} and Juna A. Kollmeier}
\affil{The Observatories of the Carnegie Institution for Science, 
    Pasadena, CA 91101}

\email{mulchaey@obs.carnegiescience.edu}
\altaffiltext{1}{Hubble Fellow/Carnegie-Princeton Fellow}

\begin{abstract}

X-ray measurements suggest the abundance of Calcium in the intracluster medium is higher than can be explained using 
favored models for core-collapse and Type Ia supernovae alone. We investigate whether the \lq\lq Calcium conundrum\rq\rq \ in the intracluster medium can be alleviated by including a 
contribution from the recently discovered subclass of supernovae known as Calcium-rich gap transients. Although the Calcium-rich gap transients make up only a 
small fraction of all supernovae events, we find that their high Calcium yields are sufficient to reproduce the X-ray measurements found for nearby 
rich clusters. We find the $\chi^{2}$  goodness-of-fit metric improves from 84 to 2 by including this new class. 
Moreover, Calcium-rich supernovae preferentially occur in the outskirts of galaxies making it easier for the nucleosynthesis products of these 
events to be incorporated in the intracluster medium via ram-pressure stripping. The discovery of a Calcium-rich gap transients in clusters and groups far
from any individual galaxy suggests supernovae associated with intracluster stars may play an important role in enriching the intracluster medium. 
Calcium-rich gap transients may also help explain anomalous Calcium abundances in many other astrophysical systems including individual stars in the Milky Way,
the halos of nearby galaxies and the circumgalactic medium.
Our work highlights the importance of considering the diversity of supernovae types and corresponding yields when modeling
 the abundance of the intracluster medium and other gas reservoirs.
\end{abstract}


\section{Introduction}

The intracluster medium (ICM) contains the majority of the baryons in clusters of galaxies. 
The presence of
heavy elements in the ICM indicates that a substantial
fraction of the diffuse gas must have passed through stars. Although the presence of iron in the 
ICM has been known for decades \citep{mitch76,serl77,mush78}, observations
with {\it ASCA}, {\it Chandra}, {\it XMM-Newton} and {\it Suzaku} have provided constraints on many additional elements. 
In particular, the abundances of O, Ne, Mg, Si, S, Ca, Ar, Ni and Fe have now been measured in    
many nearby clusters \citep{mush06,fuk98,fin00,deplaa07,wern08,bulbul12}.

Like all metals in diffuse gas, the metals in the ICM originate primarily from supernova explosions. 
Therefore, X-ray measurements of the composition of the ICM provide a means of studying the 
history of supernovae over the lifetime of the cluster. Many authors have used the abundance measurements 
to constrain the relative contribution of Type Ia and core-collapse supernovae to the 
enrichment of the intracluster medium. The abundance pattern of most clusters appears to require
a $\sim$ 30--50\% contribution from Type Ia supernovae. While a combination of Type Ia and core-collapse yields can explain the observed
abundances of most elements, recent work suggests such models underproduce the amount of Calcium required by X-ray observations
\citep{wern08,deplaa07}. 
Several solutions for the Calcium problem have been suggested in the literature including the possible underestimate of 
Calcium yields in core-collapse models \citep{wern08} or the modification of Type Ia Calcium yields based
on measurements from the Tycho supernova remnant or changing the point at which deflagration transitions to detonation \citep{deplaa07}. 

Most of the previous attempts to model the abundances of the ICM have made the simplifying assumption of uniform yields for Type Ia and core-collapse 
supernovae. However, extensive supernova searches over the last decade have revealed considerable diversity among supernovae and likely a wide range in 
yields.
Here, we investigate whether including a contribution from a new class of transients
that have nebular spectra dominated by Calcium can help explain the Calcium abundance in the ICM.

\section{Calcium-rich Gap Transients}

The existence of transients with spectra dominated by Calcium was first reported by \citet{fil03} based on observations from the 
Lick Observatory Supernova Search (LOSS; \citet{li11}). \citet{perets10} performed the first detailed study of a member of this class (SN 2005E). 
Subsequent examples of Calcium-rich supernovae have been found by both the Palomar Transient Factory \citep{mansi12} and one more by the Catalina Real Time Survey \citep{valenti13}.
Modeling the nebular spectra show that the total ejected mass of SN 2005E was very low (M$_{\it ej}$ $\sim$ 0.3 M$_{\odot}$), nearly 50\% of the ejected 
mass is in Calcium \citep{perets10}.
The total Calcium synthesized by SN 2005E is therefore a factor of 5--10 greater than that of normal Type I supernovae.

Although only a small number of Calcium-rich supernovae have been identified so far, a strong case can be made for this representing a distinctive class of objects.
All members have five distinguishing characteristics: i) peak luminosity in the gap between classic novae and supernovae; ii) rapid photometric
evolution; iii) large photospheric velocities; iv) early spectroscopic evolution into nebular phase and v) nebular spectra dominated by Calcium \citep{mansi12}.
As the peak luminosity of these physically different explosions falls in between that of novae and classic
 supernovae, these objects are referred to as Calcium-rich gap transients. In addition to the 
characteristics given above, these events appear to occur mostly in the outskirts of their hosts or in intragroup/intracluster regions, strongly suggesting 
an explosion in an old population of compact binaries. \citet{yuan13} suggest that the locations of Calcium-rich gap transients may be consistent with globular clusters. 
\citet{lyman13} show that there is no sign of in situ star formation. No model can yet explain all the characteristics of this class.

\section{Matching the Abundance Patterns of the Intracluster Medium}

In order to understand whether the large Calcium yields of the
Calcium-rich gap transients can help explain the high abundance of
Calcium seen in the ICM, we examine a series of supernova models that
include contributions from core-collapse, Type Ia and Calcium-rich gap
transients.  We compare these models to the observed ICM abundances
following the procedure outlined in \citet{deplaa07}. These ICM
measurements have the advantage of being derived from a sample of 22
clusters, while most other studies have considered one cluster at a
time. The \citet{deplaa07} dataset provides measurements for the
elements of Si, S, Ar, Ca, Fe and Ni.  

For each of the six elements under consideration, we perform a simple
linear fit to the observed abundances:
\begin{equation}
{\eta_{tot}} \sum_i \alpha_i N_{i,j} = X_{j}
\end{equation}
where $\eta_{tot}$ is the total number of supernovae, $\alpha_i$ are the fitted fraction for each of three supernova types,  
 $N_{i,j}$ is the predicted number yield of an element per supernova, and $X_{j}$ is the X-ray derived intracluster abundances for 
 each of six elements. We compute N$_{i,j}$ from mass yields as follows:
\begin{equation}
N_{i,j} = \frac{Y_{i,j}}{\mu_{j}\xi_{j}}
\end{equation}
where Y$_{i,j}$ is the predicted mass yield for each element per
supernova, $\mu_{j}$ is the mean atomic mass and $\xi_{j}$ is the
protosolar elemental abundance taken from \citet{lodders03}.  
We adopt the yields given in \citet{nomoto06} for the core-collapse supernovae.
Unfortunately, there has been very little theoretical
work on the potential yields of Calcium-rich gap transients. Therefore, we adopt the
yields given in \citet{perets10} for SN 2005E for this entire class of
Calcium-rich supernovae.

For the Type Ia yields, we first consider the WDD2 model given in
\citet{Iwa99} as this has previously been shown to provide the best fits to the X-ray data \citep{deplaa07}.
Figure 1 shows the result of taking the best-fit supernova rates given in \citet{deplaa07}, but with 10\% of the 
Type Ia supernovae assumed to be Calcium-rich gap transients. As can be seen from the Figure, including the Calcium-rich gap 
transients in the fit improves the fit ($\chi^2$ of 21 compared to 84)  for the Calcium
abundance while having little effect on the other elements.  Assuming the raw volumetric relative rates of supernovae improves the $\chi^2$ to 40 (see Table~\ref{tab:results}).

If we allow the rates of each supernova type to vary, we improve the $\chi^2$ to 2.53. Figure 2 shows this best-fit with the WDD2 model. 
The best-fit rate of Calcium-rich gap transients is 16\%  of the rate of Type Ia supernovae.
We note that volumetric rate estimates for Calcium-rich gap transients relative to Type Ia supernovae vary due to the unknown luminosity function correction and small sample size. Some recent estimates are 7$\pm$5\% \citep{perets10},  $<$20\% \citep{li11} and $>$2.3\% from the Palomar Transient Factory \citep{mansi12}.
Therefore, the best fit rate of 16\% is
consistent with current limits on rates from observational searches for these transients. 
We caution that this best-fit rate is degenerate with the Calcium yield per Ca-rich gap transient. Thus, properly accounting
for this subclass of supernovae in the modeling of the ICM abundances
may indeed provide a solution to the \lq\lq Calcium conundrum\rq\rq \ in clusters. 

Next, we consider several different models for Type Ia supernovae (see Table~\ref{tab:input}) and allow the relative fraction of each supernova type to vary. 
In Table~\ref{tab:results} we show the relative rates and derived
$\chi^2$ for each of the Type Ia models considered that yielded reasonable
fits. All models yield a $\chi^2$ that is significantly better than previous work that did not 
take into account Calcium-rich gap transients. These models have a varying degree of success at explaining the
observed abundance pattern. With the inclusion of a contribution from Calcium-rich gap transients, all of the models
reasonably fit the lower atomic number elements (see Figure 3). 
The poor $\chi^2$ values for most models are due entirely to the over or under production of nickel. Only the WDD2 and O-DDT models 
are close to reproducing the observed abundance of this element.
It is worth noting that for these two models the best-fit solutions require a higher rate for
Type Ia supernovae relative to core-collapse supernovae than is observed in the local volume surveys as has been found
by other authors.

\begin{deluxetable}{llllllllllll}
\tablewidth{7.3in}
\tablecaption{Xray abundances and model predictions.  \label{tab:input}}
\tablehead{
\colhead{\small El} &
\colhead{\small $\mu$}&
\colhead{\small $\xi$ } &
\colhead{\small [X/Fe]} &
\colhead{\small N$_{\rm Gap}$} &
\colhead{\small N$_{\rm CC}$} &
\colhead{\small N$_{\rm WDD2}$} &
\colhead{\small N$_{\rm 2003du}$} &
\colhead{\small N$_{\rm W7}$} &
\colhead{\small N$_{\rm b30}$} &
\colhead{\small N$_{\rm ODDT}$}
\\ 
\colhead{\small} &
\colhead{\small}&
\colhead{\small (10$^5$)} &
\colhead{\small } &
\colhead{\small (10$^{-8}$)} &
\colhead{\small (10$^{-8}$)} &
\colhead{\small (10$^{-8}$)} &
\colhead{\small (10$^{-8}$)} &
\colhead{\small (10$^{-8}$)} &
\colhead{\small (10$^{-8}$)} &
\colhead{\small (10$^{-8}$)}
}
\startdata
Si  &  28   &  10     &  0.68$\pm$0.12  & 0.0          & 0.4402  & 0.7333  & 0.7439   & 0.5562  & 0.1959  & 1.016  \\
S   &  32   &  4.4    &  0.6$\pm$0.05    & 0.06974 & 0.3716  & 0.8648  & 0.3347   & 0.6068  & 0.1911  & 0.8857  \\
Ar  &  36   &  1.0    &  0.40$\pm$0.03  & 0.3059   & 0.2175  & 0.6120  & \nodata  & 0.3598  & 0.1175  & 0.5386  \\
Ca  &  40  &  0.63 &  1.03$\pm$0.04  & 5.331      & 0.2679  & 0.9596  & 0.05528 & 0.4699  & 0.1425  & 0.6713  \\
Fe  &  52   &  8.4   &  1.00$\pm$0.01  & 0.8417    & 0.1940  & 1.683  & 1.764       & 1.592  & 1.129  & 1.384  \\
Ni  &  56   &  0.48  &  1.41$\pm$0.31  & 0.1063   & 0.1544  & 2.081  & 0.8511     & 4.468  & 3.759  & 2.855  \\
\enddata
\end{deluxetable}

\begin{deluxetable}{clccccc}
\tablecaption{Best Fit Results \label{tab:results}}
\tablehead{
\colhead{Rates}&
\colhead{Model} &
\colhead{$\eta_{\rm tot}$} &
\colhead{$\alpha_{\rm Gap}$} &
\colhead{$\alpha_{\rm CC}$} &
\colhead{$\alpha_{\rm Ia}$}&
\colhead{$\chi^2$}\\
\colhead{\small} &
\colhead{\small (SN\,Ia)}&
\colhead{\small} &
\colhead{\small}&
\colhead{\small} &
\colhead{\small}&
\colhead{\small}
}
\startdata
de Plaa & WDD2  & 1.34 $\times$10$^8$ & 0.00 & 0.627$\pm$0.086 & 0.373$\pm$0.012 & 84.3 \\
de Plaa $+$ Gap  & WDD2  & 1.40 $\times$10$^8$ & 0.037 & 0.630 & 0.333 & 21.1 \\ 
Volumetric & WDD2 & 1.62 $\times$10$^8$ & 0.030 & 0.701 & 0.271 & 39.9 \\ 
Best Fit & WDD2 & 0.99 $\times$10$^8$ & 0.082$\pm$0.007 & 0.407$\pm$0.094 & 0.511$\pm$0.012 & 2.53\\
\hline
Best Fit & ODDT & 0.83 $\times$10$^8$ & 0.130$\pm$0.007 & 0.098$\pm$0.100 & 0.772$\pm$0.015 & 2.21\\
Best Fit & W7   & 1.48 $\times$10$^8$ & 0.073$\pm$0.007 & 0.619$\pm$0.083 & 0.308$\pm$0.011 & 6.84\\
Best Fit & B30  & 2.04 $\times$10$^8$ & 0.054$\pm$0.007 & 0.669$\pm$0.075 & 0.277$\pm$0.013 & 9.63\\
Best Fit & 2003DU&1.70 $\times$10$^8$ & 0.078$\pm$0.007 & 0.705$\pm$0.100 & 0.219$\pm$0.010 & 9.66\\
Best Fit & CDEF & 2.27 $\times$10$^8$ & 0.050$\pm$0.007 & 0.609$\pm$0.075 & 0.341$\pm$0.018& 13.03\\
\enddata
\end{deluxetable}

\section{Discussion}

In Section 3,  we showed that simply including a contribution from the recently discovered Calcium-rich gap transients can potentially explain the high Calcium abundances reported 
for the intracluster medium. Here we explore the implications of this conclusion and what it might mean for the enrichment process in clusters of galaxies.

Although the number of Calcium-rich gap transients studied so far is small, 
they appear to reside in atypical locations for supernovae \citep{perets10,mansi12,valenti13}. 
Examples have now been found in elliptical galaxies, above the disks of edge-on galaxies and at large 
distances from isolated hosts. Moreover, all of the known examples appear to be associated with groups or clusters of galaxies. 
The remote locations of these 
events suggest that these supernovae may preferentially occur in regions of low metallicity. 
The off-galaxy locations may also work to 
maximize the contribution of Calcium-rich gap transients to the enrichment of the intracluster medium in galaxy clusters.
For example, the metals from Calcium-rich events in the outskirts of galaxies
would be more easily stripped via ram-pressure as galaxies fall into the cluster than gas concentrated near
the galaxy center. 
Similarly, Calcium produced in small groups of galaxies could easily be mixed in with the ICM when these systems get incorporated in to bigger systems. 

Given the large population of intracluster stars found in clusters some Calcium-rich supernovae may 
also occur in-situ. Supernovae with no host galaxy have indeed been found in studies of nearby clusters
\citep{galyam03,didlay10,sharon10,sand11}.
Such 
supernovae may be very efficient at polluting the intracluster medium
as their metals can be directly deposited into
the cluster without needing to escape the confining potential of a galaxy
\citep{gal00,zaritsky04,siv09,rasmussen10}. 
The recent discovery of a Calcium-rich gap transient in the Coma
cluster far away from any galaxies (Kasliwal et al. 2013, in prep.) supports the idea that some Calcium-rich gap transients are indeed associated with intracluster stars.
Future transient surveys should provide constraints on the importance of intracluster Calcium-rich gap transients.

Calcium-rich gap transients may also play an important role in enriching gas in other astrophysical systems. Although the uncertainties on the measurements are large,
the Calcium abundance of the intragroup medium in some groups of galaxies appears to be even higher than that in rich clusters \citep{grange11}. This might be expected
given that most of the Calcium-rich gap transients discovered to date occur in small groups and many of the stripping mechanisms at play in clusters are also 
effective in lower mass systems \citep{rasmussen06,kawata08}. The possible association of Calcium-rich gap transients with low metallicity regions suggests that these 
transients may also be an importance source of enrichment in the halos of individual galaxies. This could help account for Calcium-overabundant stars in the 
halo of the Milky Way \citep{andy95,lai09} and the large amount of Calcium in the circumgalactic medium \citep{zhu13}. 
Furthermore, the tendency for the Calcium abundance to track Iron rather than Oxygen in early-type galaxies \citep{conroy13} 
suggests most of the Calcium in these galaxies is not produced by the core-collapse supernovae. Rather, the source of Calcium is likely dominated by Type Ia and Calcium-rich
gap transients as we find for rich clusters (see Figure 2).

While our work suggests the potential importance of Calcium-rich gap transients on the enrichment of the 
intracluster medium, further study will be required to reach firm conclusions.
For example, for our calculations we have relied on the yield 
estimates for a single supernova for the entire class of Calcium-rich gap transients.
High quality nebular spectra are now available for several
Calcium-rich gap transients and more detailed theoretical modeling of
these spectra would provide a more representative average yield of
Calcium per transient. Second, as with previous work, the
core-collapse yields here include the contribution of both Type II and
Type Ibc supernovae. Both the rates and the yields of each of these
sub-types is different and it may be more accurate to consider their
contributions separately.  Third, the yields as determined by
theoretical methods and nebular spectrum modeling methods are not
consistent for Type Ia supernovae. More work must be done to
understand these differences and their impact on the yield estimates.
Finally, with the recent proliferation of all-sky transient surveys, multiple new classes of elusive explosions have been uncovered
(see e.g. brief review in \citealt{kasliwal12b} and references therein). Yield estimates for each of these new classes should
be undertaken to assess their contribution to the ICM.  
While exploring these effects is beyond the scope of this {\it
  Letter}, we hope that our simple calculations will provide
motivation for further theoretical modeling for all supernova types.

The need for more detailed supernovae modeling is further justified by upcoming X-ray missions like ASTRO-H \citep{tak10}, 
which will allow spatially resolved high-resolution X-ray spectroscopy for clusters for the first time. Detailed maps of the spatial 
variation of elements may provide important clues about the stars responsible for enriching the intracluster medium.


\acknowledgments

We are grateful to Jelle de Plaa and Esra Bulbul for providing additional insight on their 
previous work and thank Avishay Gal-Yam, Melissa Graham and Paolo Mazzali for helpful discussions.
This work was inspired by presentations at the conference Energetic Astronomy: Richard Mushotzky at 65. We thank
Chris Reynolds for his help organizing the conference. 



{\it Facilities:} \facility{XMM-Newton}, \facility{CXO}


\begin{figure}
\epsscale{.80}
\plotone{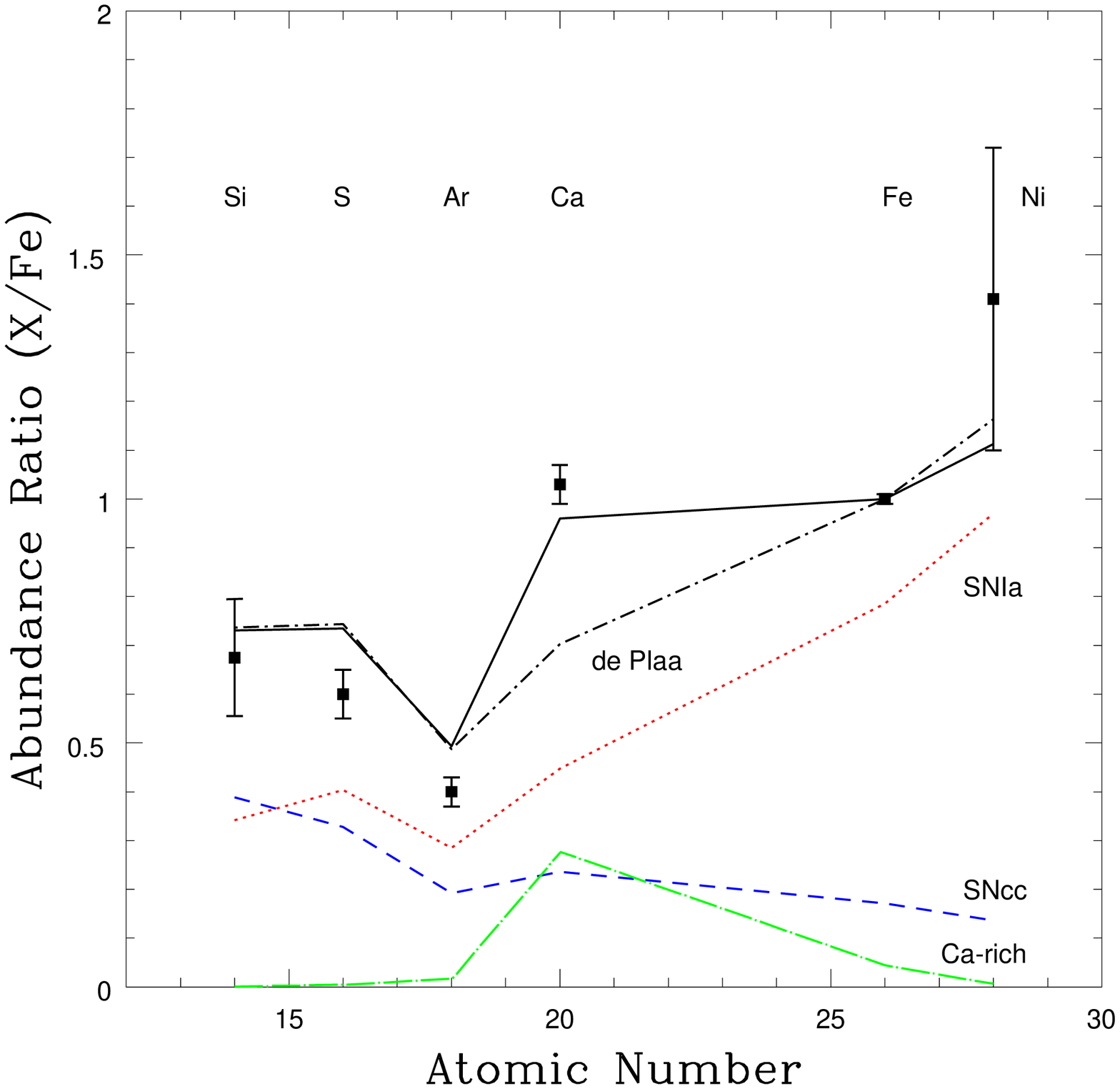}
\caption{Abundance ratios relative to iron versus atomic numbers for the cluster sample given in \citet{deplaa07}. We show the contributions of the three types of supernovae (Type Ia, 
core-collapse and Calcium-rich gap transients) assuming the best-fit supernovae rates given in \citet{deplaa07}, but with 10\% of the Type Ia's assumed to be 
Calcium-rich gap transients. The solid black line shows the total abundance ratio. The best-fit found by \citet{deplaa07} considering only the contribution from 
Type Ia's and core-collapse supernovae is shown by the black dot-short dash line. The inclusion of a contribution from the Calcium-rich gap transients increases the overall Calcium abundance, 
but has little impact on the other elements.} 
\end{figure}

\clearpage

\begin{figure}
\epsscale{.80}
\plotone{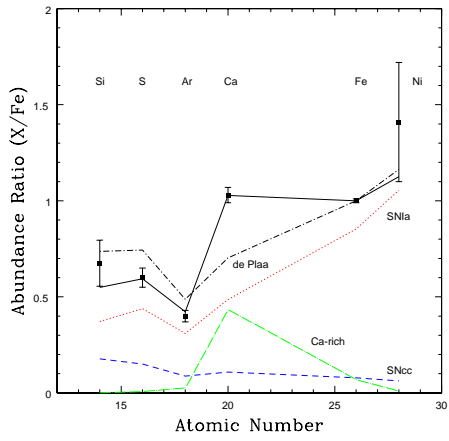}
\caption{Same as Figure 1, but with the rates of each supernovae type allowed to vary.}
\end{figure}

\clearpage

\begin{figure}
\epsscale{.80}
\plotone{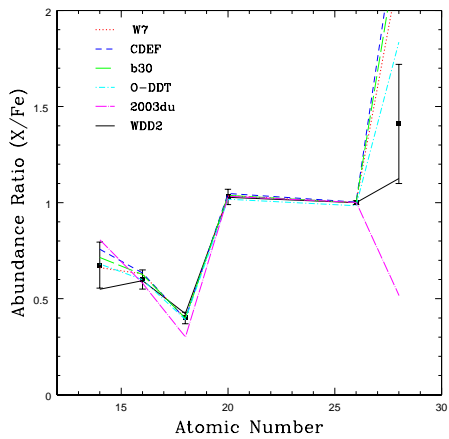}
\caption{Fits to the data given in Figure 1 using the yields from a number of different Type Ia models. The core-collapse and Calcium-rich gap transients yields are the same as those 
used in Figure 1, but the rates for each supernovae type were allowed to vary. While all of the Type Ia models can reproduce the lower atomic
number elements, most over or under produce Nickel.}
\end{figure}

\end{document}